\documentclass[article]{aa}
\usepackage[dvips]{graphicx}
\usepackage{txfonts}
\usepackage{natbib}
\bibpunct{(}{)}{;}{a}{}{,}

\begin{document}
\title{RXTE and XMM observations of intermediate polar candidates}

\author{
  O. W. Butters\inst{1}
\and  
  A. J. Norton\inst{2}
\and
  K. Mukai\inst{3}
\and
  J. A. Tomsick\inst{4} 
}

\institute{
  Department of Physics and Astronomy, University of Leicester,
  Leicester, LE1 7RH, UK\\
  \email{oliver.butters@star.le.ac.uk}
\and
  Department of Physics and Astronomy, The Open University, Walton
  Hall, Milton Keynes MK7 6AA, UK
\and
 CRESST and X-ray Astrophysics Laboratory NASA/GSFC, Greenbelt,
        MD 20771, USA {\it and} Department of Physics, University of Maryland,
        Baltimore county, 1000 Hilltop Circle, Baltimore, MD 21250, USA 
\and
 Space Sciences Laboratory, 7 Gauss Way, University of California,
 Berkeley, CA 94720-7450, USA
}

\authorrunning{Butters et al.}

\date{Accepted 2010 ???;
      Received  2010 ???;
      in original form 2010 ???}

\abstract
{}
{To determine the credentials of nine candidate intermediate
polars in order to confirm whether or not they are magnetic cataclysmic variables.}
{Frequency analysis of \emph{RXTE} and \emph{XMM} data was used to search for temporal
variations which could be associated with the spin period of the magnetic white
dwarf. X-ray spectral analysis was carried out to characterise the emission
and absorption properties of each target.}
{The hard X-ray light curve of V2069~Cyg shows a pulse period of 743.2~s, and its spectrum is fit 
by an absorbed bremsstrahlung model with an iron line, confirming this to be a genuine intermediate polar. 
The hard X-ray light curve of the previously confirmed intermediate polar IGR J00234+6141 is 
shown to be consistent with the previous low energy X-ray detection of a 563.5~s pulse period.
The likely polar IGR J14536--5522 shows no coherent modulation at the previously identified period
of 3.1~hr, but does exhibit a clear signal at periods likely to be harmonically related to it. 
Whilst our \emph{RXTE} observations of RX J0153.3+7447, Swift J061223.0+701243.9, V436 Car and DD Cir
are largely too faint to give any definitive results, the observation
of IGR J16167--4957 and V2487~Oph show some characteristics of
intermediate polars and these objects remain good candidates.}
{We confirmed one new hard X-ray selected intermediate polar from our sample, V2069 Cyg.}


\maketitle

\section{Introduction}

Intermediate polars (IPs) are a sub-class of cataclysmic variables
(CVs) which have a strong magnetic
field on the white dwarf. The strength of the magnetic field is
generally believed to be somewhere between that of the non-magnetic CVs and
their sister class the polars. This gives an expected range of a few
MG to a few tens of MG at the white dwarf surface.

The magnetic field is large enough to dramatically alter the accretion flow,
but not large enough to synchronise the spin and orbital periods. This
gives rise to flux variations in the optical and X-ray bands pulsed at the
spin period (as well as harmonics and beat periods). For a good overview of CVs
see e.g. \cite{warner95}.

Their is no formal definition of what characteristics an IP
must have to be definitively classed as a genuine member. Common features
include coherent optical and X-ray modulations at the white dwarf spin period (and/or at the 
the beat period between the white dwarf spin and orbital periods) and an X-ray spectrum 
fit by an absorbed bremsstrahlung model, usually with a strong fluorescent iron line 
at 6.4~keV. The population, and classifications used here are those of Koji
Mukai\footnote{http://asd.gsfc.nasa.gov/Koji.Mukai/iphome/iphome.html
  (IP catalogue version 2009a)}, and therefore we consider there to be
34 confirmed IPs. 

The hard X-ray \emph{INTEGRAL} mission has found many of the existing population
of IPs in its survey \citep{barlow06}. This has raised the question of whether the
existing population is biased towards being soft X-ray selected. With this in
mind we started a programme to identify and characterise hard X-ray selected
candidate IPs  \citep{butters07,butters08,butters09}. The
work presented here is the final stage of that process and presents an investigation
of nine further candidate IPs with \emph{RXTE}. Most of these objects 
(i.e. IGR J00234+6141, Swift J061223.0+701243.9, IGR J14536--5522, IGR
J16167--4957, V2487~Oph 
and V2069 Cyg) have been selected on the basis of their hard X-ray emission. However, we 
also include a few IP candidates (i.e. RX J0153.3+7446, V436 Car and DD Cir) which have had 
no previous reports of hard X-ray emission, for comparison. Each target has previously been 
observed in the optical and classed as a likely IP, with usually the X-ray variability 
missing in order to definitively classify it.

\section{Observations and data reduction}

Data were obtained from the \emph{RXTE} satellite \citep{bradt93} with the PCA 
instrument in cycles 12 and 13. Initial data reduction was done with the standard {\sc
  ftools}. Only the top layer of each PCU was included in the measurements and the time resolution of
the data was 16~s. Background subtracted light curves were constructed in four 
energy bands: 2--4~keV, 4--6~keV, 6--10~keV and 10--20~keV, as well as 
a combined 2--10~keV band for maximum signal-to-noise. Each of these
were then searched for periodicities using the {\sc clean} algorithm of
\citet{lehto97}, which iteratively de-convolves the window function from
the data. Where a period was found the data were folded at this period
and an estimate of modulation depth made.

A mean X-ray spectrum was also extracted, and two simple spectral
models fitted to it: a bremsstrahlung and a power law, each absorbed by a 
single column density of neutral hydrogen. In each case the
energy range 5.5--7.5~keV was ignored to achieve a fit to the bulk of
the data, then this range was added back in to fit any excess that
might be present due to iron line emission. If the
fitted spectrum could not fit the Galactic column density well then an
interpolated value from \citet{dickey90} was calculated using the
HEASARC $n_H$ web
interface\footnote{http://heasarc.gsfc.nasa.gov/cgi-bin/Tools/w3nh/w3nh.pl},
denoted D\&L in the text.  

\emph{RXTE} has a large field of view, therefore for each target the \emph{ROSAT} Bright
Source Catalogue was searched for nearby sources. Any sources present
had their \emph{RXTE} count rate estimated with
webPIMMS\footnote{www.ledas.ac.uk} and scaled according to their
position on the detector.

In the case of IGR J16167$-$4957 we also present \emph{XMM} data
alongside the \emph{RXTE} data. The {\em XMM} data was taken from an
observation made in August 2006 (ObsID 0402920101). The EPIC pn and
MOS instruments \citep{struder01,turner01} accumulated photons in
``small window'' mode with the medium filter. In addition to the {\em
  XMM} data, we obtained the calibration files indicated as necessary
for this observation by the on-line software tool {\ttfamily
  cifbuild}. We used the Science Analysis Software (SAS-10.0.0)
package to reprocess the pn, MOS1, and MOS2 data using {\ttfamily
  epproc} and {\ttfamily emproc}, yielding photon event lists for the
three instruments. Although proton flares sometimes cause portions of
{\em XMM} observations to have very high backgrounds, we did not find
evidence for proton flares, and we were able to use the full exposure
time. 

We used the SAS tool {\ttfamily xmmselect} to produce {\em XMM} energy
spectra for IGR~J16167--4957.  We used standard event filtering and
included photons within an aperture with a $30^{\prime\prime}$
radius. We took background spectra from the nearest source-free region
of the CCD. For the MOS detectors, we used the 0.1--10~keV bandpass,
and for the pn detector, we used the 0.2--12~keV bandpass. 

The observing log can be found in Table~\ref{observing_log}, and the
targets are now presented in order of R.A.

\begin{table*}
\centering
\caption{Observing log. Targets are ordered by R.A.}
\label{observing_log}
\centering
\begin{tabular}{llllllll}
\hline
\hline
Target                   & R.A.        & Declination   & Abbreviated & Obs.        & Start time        & End time          & Good time\\
                         &             &               & name$^*$    &             &                   &                   & \\
                         &             &               &             &             & (UTC)             & (UTC)             & (s)   \\
\hline
IGR J00234+6141          & 00 22 57.64 & $+$61 41 07.6 & J0023       & \emph{RXTE} & 17:29:50 01/03/09 & 16:20:17 03/03/09 & 41\,776 \\
RX J0153.3+7446          & 01 53 21.01 & $+$74 46 21.9 & J0153       & \emph{RXTE} & 00:15:44 20/04/09 & 11:14:17 21/04/09 & 41\,376 \\
Swift J061223.0+701243.9 & 06 12 22.6  & $+$70 12 43.4 & J0612       & \emph{RXTE} & 00:24:37 30/03/09 & 06:16:17 31/03/09 & 39\,216 \\
V436 Car                 & 07 44 57.93 & $-$52 57 13.8 & --          & \emph{RXTE} & 15:45:44 08/03/09 & 13:35:17 10/03/09 & 27\,456 \\
DD Cir                   & 14 23 23.41 & $-$69 08 45.2 & --          & \emph{RXTE} & 07:18:01 25/12/09 & 11:10:17 26/12/09 & 49\,312 \\
IGR J14536$-$5522        & 14 53 41.06 & $-$55 21 38.7 & J1453       & \emph{RXTE} & 03:39:12 24/11/07 & 04:00:32 25/11/07 & 31\,320 \\
IGR J16167$-$4957        & 16 16 37.20 & $-$49 58 47.5 & J1616       & \emph{XMM}  & 16:11:15 17/08/06 & 22:25:40 17/08/06 & 22\,465 \\
IGR J16167$-$4957        & 16 16 37.20 & $-$49 58 47.5 & J1616       & \emph{RXTE} & 10:21:59 17/12/09 & 14:13:17 18/12/09 & 34\,864 \\
V2487 Oph                & 17 31 59.8  & $-$19 13 56   & --          & \emph{RXTE} & 22:20:18 20/08/10 & 14:29:17 22/08/10 & 42\,312 \\
V2069 Cyg                & 21 23 44.82 & $+$42 18 01.7 & --          & \emph{RXTE} & 04:05:49 07/12/09 & 11:22:17 08/12/09 & 37\,472 \\
\hline
\hline
\multicolumn{8}{l}{$^*$ Used henceforth.}\\
\end{tabular}
\end{table*}


\section{The candidates}

\subsection{IGR J00234+6141}

This is classified as an `ironclad' IP (according to the IP catalogue 
version 2009a) with both an established spin and orbital period in the 
optical ($P_{spin}=563.5$~s and $P_{orb}=14\,520$~s respectively 
\citep{bonnetbidaud07}). Further to this \citet{anzolin09} report 
\emph{XMM} observations which clearly show spin modulated X-ray variability 
below 2~keV, but not above it. No hard X-ray detections of 
the white dwarf spin period have been reported. \citet{tomsick08}
report \emph{Chandra} spectra over a similar energy range.

\subsubsection{Observations and results}

\begin{figure}
  \includegraphics[width=\hsize]{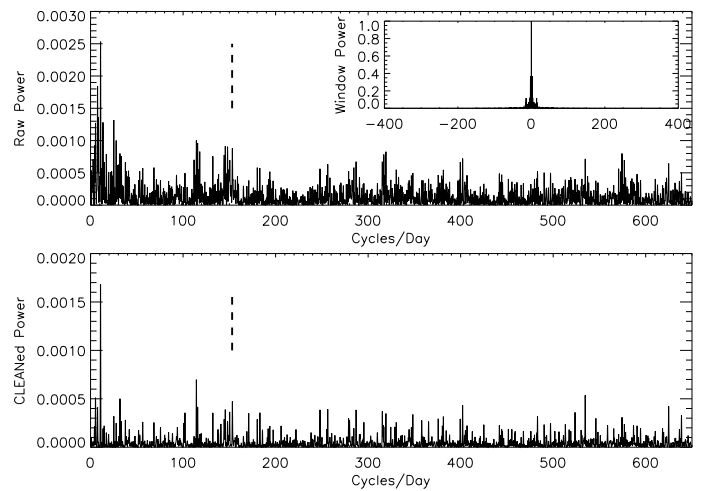}
  \caption{The {\sc clean}ed power spectrum of the 2--10~keV light curve of J0023. 
  The upper plot shows the raw power spectrum with the window function insert; the lower
  plot shows the deconvolved ({\sc clean}ed) power spectrum. The
  dashed line is at 153 cycles~day\textsuperscript{-1} (563.5~s).}
  \label{J0023_cleaned}
\end{figure}

J0023 was observed over two consecutive days with a total of 41\,776~s
good time in PCU2 (see Table~\ref{observing_log}). The 2--10~keV energy
band raw count rate varied between 2.0--6.1~ct~s$^{-1}$ and the
background count rate was 2.9--4.0~ct~s$^{-1}$. The mean background
subtracted count rate in the 2--10~keV band was 0.7~ct~s$^{-1}$.
The {\sc clean}ed power spectrum of the 2--10~keV \emph{RXTE} light curve 
(Fig.~\ref{J0023_cleaned}) has a peak at
153 cycles~day$^{-1}$ (where the 563.5~s spin period would be expected
to be seen), however, this is in a forest of peaks, each with a
similar power to the `noise' seen elsewhere. The peak at $\sim$10
cycles~day$^{-1}$ is too close to the length of some of the data sets
to be considered legitimate here.

\begin{figure}
  \includegraphics[width=\hsize]{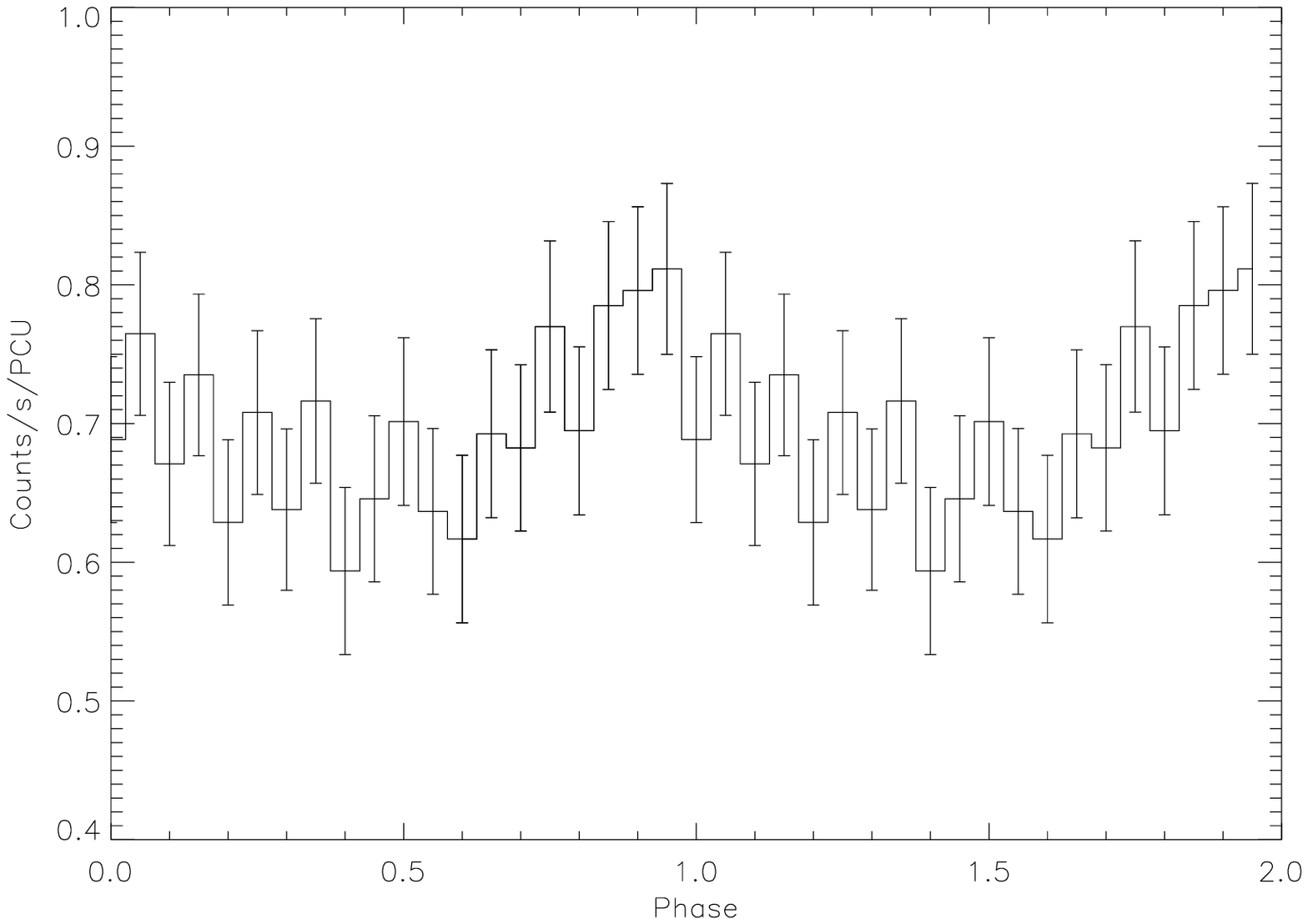}
  \caption{The 2--10~keV background subtracted light curve of J0023 folded
    at 563.5~s with an arbitrary zero point. Two cycles are shown.}
  \label{J0023_2_10_folded}
\end{figure}

The 2--10~keV light curve folded at the previously identified spin period is 
shown in Fig.~\ref{J0023_2_10_folded} where a single peaked modulation is seen.
Folding the light curve in different energy bands gives a trend of increasing 
modulation depth with increasing energy (summarised in Table~\ref{J0023_modulation_depth}). 
Folding the data at the 14\,520 s orbital period gives an indication of
a double peaked modulation, however this is by no means conclusive.

\begin{table}
\centering
\caption{Modulation depths of the J0023 light curve folded at 563.5~s.}
\begin{tabular}{lll}
\hline
\hline
Energy & Mean count    & Modulation depth\\
(keV)  & (ct~s$^{-1}$) & (\%)\\
\hline
2--4   & $0.19\pm0.01$ &  $4.7\pm5.0$\\
4--6   & $0.25\pm0.01$ &  $5.9\pm4.3$\\
6--10  & $0.26\pm0.01$ & $15.8\pm4.8$\\
10--20 & $0.09\pm0.01$ & $48.4\pm16.4$\\
\hline
\hline
\label{J0023_modulation_depth}
\end{tabular}
\end{table}

\begin{figure}
\resizebox{\hsize}{!}{\rotatebox{-90}{\includegraphics[width=\hsize]{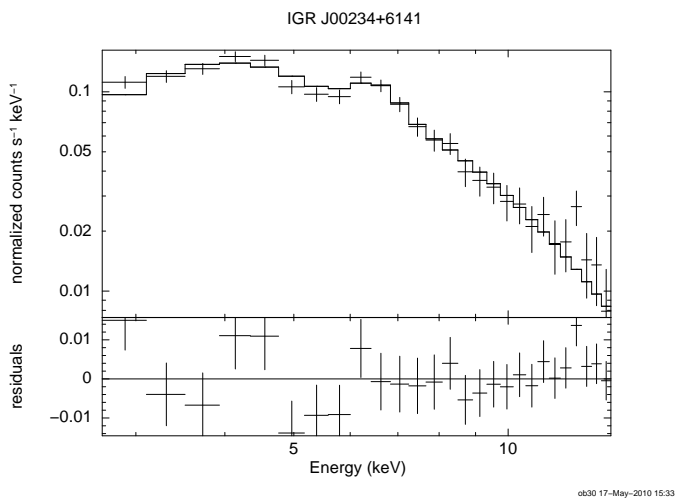}}}
  \caption{Bremsstrahlung model fit to the 2.5--12~keV spectrum of J0023.}
  \label{J0023_spectrum}
\end{figure}

The parameters of the spectral models used to fit the J0023 X-ray spectrum 
are shown in Table~\ref{spectral_fits}. In both cases the column density had to be
pegged to a lower limit. An iron emission line is clearly present and the spectrum 
is hard, as shown in Fig.~\ref{J0023_spectrum}.

\subsubsection{Discussion}

The trend of increasing modulation depth with energy is the opposite to what is commonly seen 
in other IPs, where photoelectric  absorption is responsible for much of the modulation. 
However, the modulation depth of the folded light curves in the 2--4 and 4--6~keV bands is close to being
consistent with zero. This is in good agreement with \citet{anzolin09}
who found no modulation above 2~keV. We note that there is some indication of
significant modulation above 6~keV in our data and conclude that this is at least
consistent with the previously detected white dwarf spin period.

The spectral properties are quite different from those of
\citet{anzolin09} and \citet{tomsick08}, and are hard to 
explain. However, there is a source near by that may have a comparable count rate to that measured
here and which is a likely contaminant in these \emph{RXTE} data. 

We conclude that our \emph{RXTE} data are consistent with the previous IP classification of 
J0023, but are unable to shed much further light on its properties.

\begin{table*}
\centering
\caption{Spectral fit properties. Uncertainties are given at the 90\% confidence level.}
\centering
\begin{tabular}{llllllllll}
\hline
\hline
Target          & Model$^+$ & $n_H$              & kT           & $\Gamma$    & E           & $\sigma_E$  & EW   & $\chi_r^2$ & Flux (2--10 keV)\\
                &           & 10$^{22}$~cm$^{-2}$ & keV          &              & keV         & keV         & keV  &            & 10$^{-11}$~ergs~cm$^{-2}$~s$^{-1}$\\
\hline
J0023           & B         & 0.733$^{*}$        & $10.8\pm2.1$ & --           & $6.5\pm0.2$ & $0\pm0.3$   & 0.5  & 1.2        & 0.77\\
J0023           & P         & 0.733$^{*}$        & --           & $1.9\pm0.1$  & $6.6\pm0.2$ & $0\pm0.4$   & 0.6  & 0.94       & 0.79\\



V436 Car        & B         & 0.14$^{*}$         & $11.7\pm4.8$ & --            & --          & --          & --  & 1.3        & 0.48\\
V436 Car        & P         & 0.14$^{*}$         & --           & $1.9\pm0.2$   & --          & --          & --  & 1.1        & 0.50\\


J1453           & B         & 0.524$^{*}$        & 25$\pm$1     & --           & 6.4$\pm$0.1 & 0.3$\pm$0.1 & 0.4  & 1.1        & 4.1\\
J1453           & P         & 0.8$\pm$0.8        & --           & 1.6$\pm$0.1  & 6.3$\pm$0.1 & 0.5$\pm$0.1 & 0.5  & 0.9        & 4.1\\

J1616$^{\#}$     & B         & 2.16$^{*}$         & 17           & --            & 6.4         & 0           &      & 3.5       & 5.7\\
J1616           & P         & 2.16$^{*}$         & --           & 1.75$\pm$0.02 & 6.4$\pm$0.1 & 0.1$\pm$0.1 & 0.6  & 1.2        & 5.6\\

V2487 Oph$^{\#}$ & B         & 0.2$^*$            & 10.4         & --            & 6.4         & 0           & 0.5  & 2.7       & 1.2\\
V2487 Oph       & P         & 0.2$^*$            & --           & 1.96$\pm$0.07 & 6.4$\pm$0.1 & 0.2$\pm$0.2 & 0.6  & 1.6        & 1.2\\

V2069 Cyg       & B         & 0.362$^{*}$        & 50$\pm$20    & --            & 6.3$\pm$0.1 & 0$\pm$0.3   & 0.5  & 0.5        & 1.2\\
V2069 Cyg       & P         & 0.362$^{*}$        & --           & 1.4$\pm$0.1   & 6.3$\pm$0.1 & 0.2$\pm$0.2 & 0.5  & 0.5        & 1.2\\

\hline
\hline
\multicolumn{9}{l}{+ - B=Bremsstrahlung, P=Power law.}\\
\multicolumn{9}{l}{* - Pegged to a lower limit derived from D\&L.}\\
\multicolumn{9}{l}{\# - No uncertainty estimate as the fit was poor.}\\
\label{spectral_fits}
\end{tabular}
\end{table*}

\subsection{RX J0153.3+7446}

J0153 was discovered in the \emph{ROSAT} all-sky survey and a spin period of 1414~s 
was suggested \citep{haberl95}. An orbital period of 14\,183 s (3.9396
hr) has been proposed (private communication cited in \citet{downes01})
but never published. \citet{norton06} carried out optical photometry
and found strong modulation at a period of 2333$\pm$5~s. They also reanalysed 
the \emph{ROSAT} data and concluded that the true spin period is probably 
1974$\pm$30~s with the 2333~s optical period then being the beat with the 
3.9396 hr orbital period.

J0153 was observed with \emph{RXTE} over two consecutive days with a total of 41\,376~s
good time in PCU2 (see Table~\ref{observing_log}). The 2--10~keV energy
band raw count rate varied between 1.9--5.8 ct~s$^{-1}$ and the background was 
2.8--4.1 ct~s$^{-1}$. The mean background subtracted count rate in this 
energy band was 0.18 ct~s$^{-1}$.

\begin{figure}
  \includegraphics[width=\hsize]{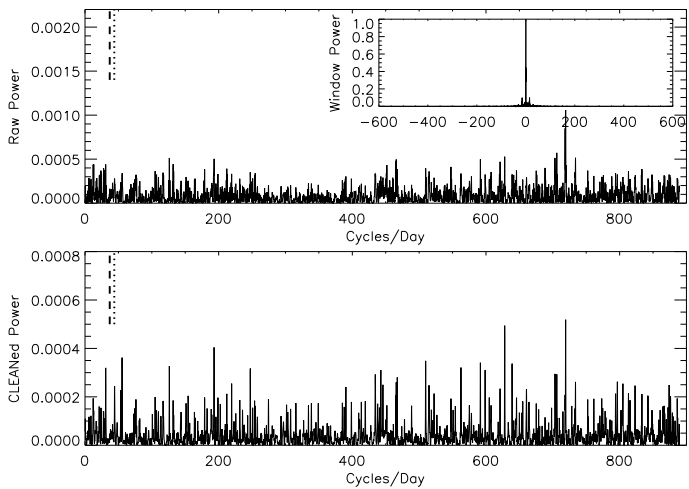}
  \caption{The {\sc clean}ed power spectrum of the 2--10~keV light curve of J0153. 
  The upper plot shows the raw power spectrum with the window function insert; the lower
  plot shows the deconvolved ({\sc clean}ed) power spectrum. The
  dashed line is at 37 cycles~day\textsuperscript{-1} (2333~s), the dotted line at
  44 cycles~day\textsuperscript{-1} (1974~s).}
  \label{J0153_cleaned}
\end{figure}

 Fig.~\ref{J0153_cleaned} shows the {\sc clean}ed power spectrum of the 2--10~keV
background subtracted light curve. No significant peaks are seen, but a peak \textit{is}
present (albeit a small one) at 1974~s (43.8 cycles
day$^{-1}$). However, folding the data at the 1974~s and 2333~s periods yields
no significant coherent modulation in either case at each energy band.

Unfortunately, the flux in the background subtracted X-ray spectrum is too faint 
to achieve a  meaningful fit to the data. Furthermore, there is another X-ray source 
near by that is likely to have a comparable count rate to that measured here, which 
\citet{veron06} classify as a quasar.

The detection of J0153 in this RXTE observation is therefore sadly too faint to form
any definitive conclusion as to its nature.

\subsection{Swift J061223.0+701243.9}

Very little is known about J0612. It was discovered inside the field
of view of a \emph{Swift} BAT trigger \citep{grupe06_gcn} and was seen 
to brighten then fade over a period of 58~ks while \emph{Swift} was 
observing it. It was later observed again with \emph{Swift} for $\sim$8~ks 
to measure a spectrum \citep{grupe06_atel}. The spectrum was modelled by 
an absorbed power law with $\Gamma=1.61\pm0.24$ and $N_H=(1.8\pm0.8)\times
10^{21}$~cm$^{-2}$ and the flux was $3.20 \times
10^{-12}$~ergs~cm$^{-2}$~s$^{-1}$ in the 0.3--10~keV
band. \citet{grupe06_atel} also report low resolution spectra from the
Hobby-Eberly-Telescope at McDonald Observatory. The optical spectrum
clearly showed strong hydrogen and helium lines, identifying this
target as either a low mass X-ray binary or a CV.

J0612 was observed with \emph{RXTE} over two consecutive days with a total of 39\,216~s
from PCU2 (see Table~\ref{observing_log}). The 2--10~keV raw count rate
varied between 1.9--5.3~ct~s$^{-1}$ and the background subtracted count
rate had a mean of close to 0.0~ct~s$^{-1}$. The 
background may therefore be over-subtracted, but in any case the source
is at the limit of detectability in this observation.

\begin{figure}
  \includegraphics[width=\hsize]{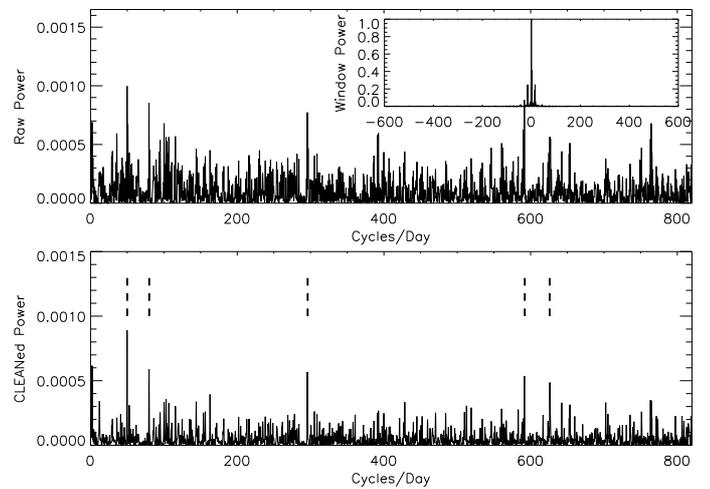}
  \caption{The {\sc clean}ed power spectrum of the 2--10~keV light curve of J0612. 
  The upper plot shows the raw power spectrum with the window function insert; the lower
  plot shows the deconvolved ({\sc clean}ed) power spectrum. The
  dashed lines represent the five main peaks (see text).}
  \label{J0612_cleaned}
\end{figure}

\begin{figure}
  \includegraphics[width=\hsize]{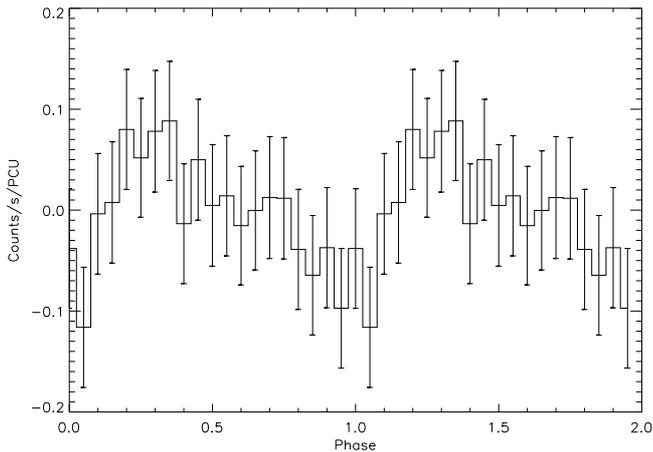}
  \caption{The 2--10~keV background subtracted light curve of J0612 folded
    at 292.3~s with an arbitrary zero point. Two cycles are shown.}
  \label{J0612_folded_at_292p3}
\end{figure}

Fig.~\ref{J0612_cleaned} shows the {\sc clean}ed power spectrum of the 2--10~keV
background subtracted light curve. Five peaks stand out in the plot, 50 cycles~day$^{-1}$
($1734\pm9$~s), 80 cycles~day$^{-1}$ ($1083\pm4$~s), 296 cycles~day$^{-1}$
($292.3\pm0.3$~s), 592 cycles~day$^{-1}$ ($146.0\pm0.1$~s), and 626
cycles~day$^{-1}$ ($138.0\pm0.1$~s). The 2--10~keV light curve folded at 
the one of these that is most typical of an IP spin period (i.e. 292.3~s)
is shown in Fig.~\ref{J0612_folded_at_292p3}. A coherent modulation is seen,
but the error bars are large and the profile is consistent with zero modulation
(and indeed zero flux). 

As with J0153, the background subtracted spectrum is too faint to obtain a meaningful
fit. An uncategorised X-ray source is also in the \emph{RXTE} field of view, and may
contribute a comparable count rate to the target. We also note that
J0162 is not found in the \emph{ROSAT} all sky survey, so is likely to be a transient object. Sadly,
given its faintness in this observation, we cannot comment further on its nature.

\subsection{V436 Car}

V436 Car (RX J0744.9$-$5257) was identified as a CV by \citet{motch96}
in the \emph{ROSAT} all-sky survey. \citet{ramsay98} carried out optical and
X-ray analysis of the object, concluding it was a likely IP. They found a
probable orbital period of 3.60~h from optical spectrometry along with
several other candidate periods from aliases and their
photometry. \citet{woudt02} carried out high speed photometry of V436
Car, concluding the true period was in fact 4.207~h, an alias of the
period seen by \citet{ramsay98}.

V436 Car was observed with \emph{RXTE} over two consecutive days with a total of 27\,456
s good time in PCU2 (see Table~\ref{observing_log}). The 2--10~keV
energy band raw count rate varied between 2.3--7.3~ct~s$^{-1}$, and
the background count rate was 0--4.2~ct~s$^{-1}$. The mean background
subtracted count rate in the 2--10~keV band was 0.43~ct~s$^{-1}$.
The data is stable over the run with no obvious flaring events. 

\begin{figure}
  \includegraphics[width=\hsize]{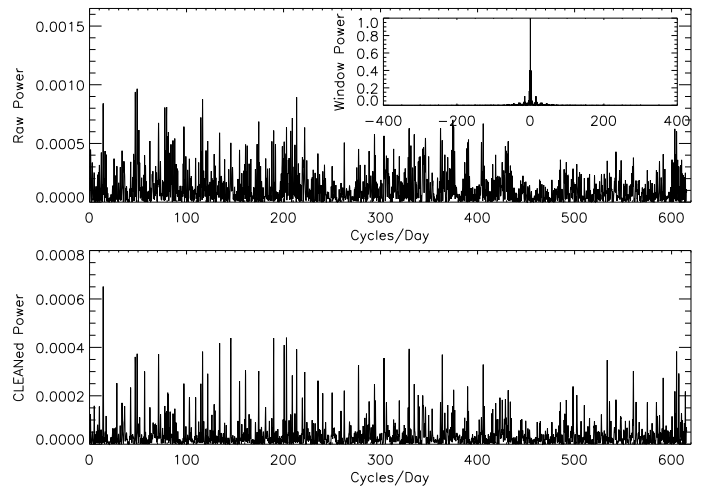}
  \caption{The {\sc clean}ed power spectrum of the 2--10~keV light curve of V436 Car. 
  The upper plot shows the raw power spectrum with the window function insert; the lower
  plot shows the deconvolved ({\sc clean}ed) power spectrum.}
  \label{V436_Car_cleaned}
\end{figure}

Fig.~\ref{V436_Car_cleaned} shows the {\sc clean}ed power spectrum of the 2--10~keV
background subtracted light curve. The largest peak is at $\sim 14$ cycles day$^{-1}$ 
(1.706$\pm$0.025 hr). None of the orbital periods found by \citet{ramsay98} or
\citet{woudt02} have been found here and no significant coherent modulation that is
commensurate with a spin period is seen here either. This agrees with
the lack of detection in the previous studies in both the optical and
X-ray.

The X-ray spectrum of V436 Car is relatively poorly fit with both models 
(see Table~\ref{spectral_fits}) and shows no sign of an iron emission line. 
There is another source in the \emph{RXTE} field of view, but it is sufficiently faint
and far from the centre that it should not affect this result.

Given the lack of coherent modulation and absence of an iron line in the spectrum, it is
unlikely that V436~Car is an IP.

\subsection{DD Cir}

DD Cir was first discovered as a nova with an apparent magnitude of 7.7
\citep{liller99}.  \citet{woudt03} subsequently carried out high speed photometry
using the 74-inch telescope at the SAAO and their run yielded an average
magnitude (not standardised) of $\sim$20. They found DD~Cir to be an eclipsing 
system with an orbital period of 2.339~hr, and concluded it had an inclination of
$79^{\circ}$. Further to this, \citet{woudt03} found distinct
periodic signals in three individual nights' FTs which were consistent
within errors, at $\sim670$~s.

Data was collected with \emph{RXTE} over just over one day (see
Table~\ref{observing_log}). The total good time on target was 49\,212~s
in PCU2. The raw target count rate varied between 3.5 and 8.3
ct~s$^{-1}$ and the background was 4.5--6.7~ct~s$^{-1}$. The mean background
subtracted rate was 0.1  ct~s$^{-1}$ in the 2--10~keV band.

\begin{figure}
  \includegraphics[width=\hsize]{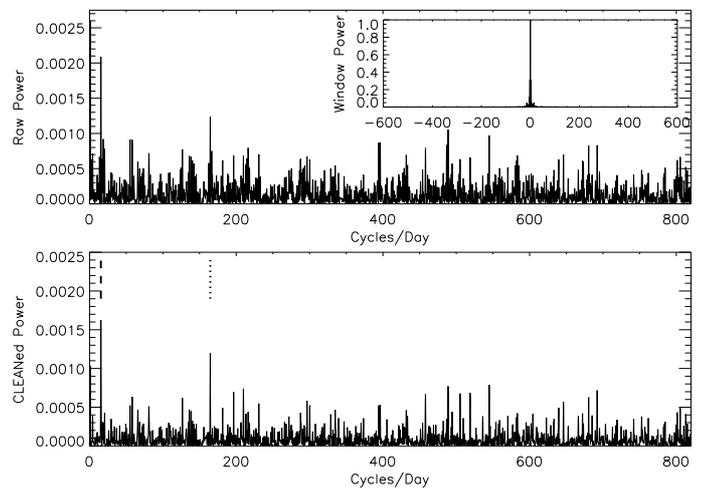}
  \caption{The {\sc clean}ed power spectrum of the 2--10~keV light curve of DD Cir. 
  The upper plot shows the raw power spectrum with the window function insert; the lower
  plot shows the deconvolved ({\sc clean}ed) power spectrum. The
  dashed line is at 15.5 cycles~day$^{-1}$, the dotted line at 165 cycles~day$^{-1}$.}
  \label{DD_Cir_cleaned}
\end{figure}

\begin{figure}
  \includegraphics[width=\hsize]{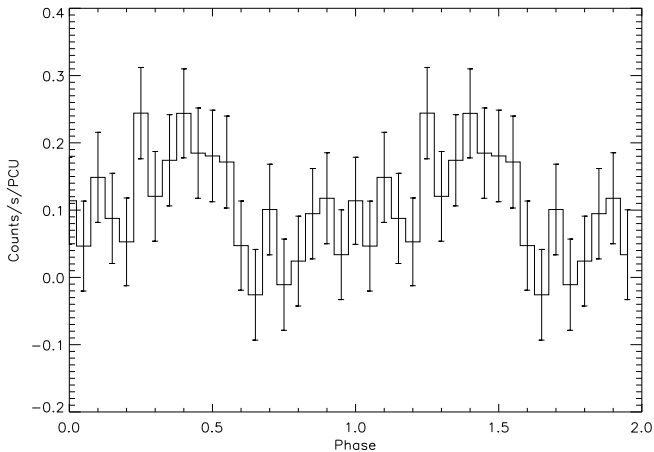}
  \caption{The background subtracted 2--10~kev light curve of DD
    Cir folded at 525~s. The zero point on the time axis is the time of the first
    observation.}
  \label{DD_Cir_folded}
\end{figure}

Fig.~\ref{DD_Cir_cleaned} shows the {\sc clean}ed power spectrum of the background
subtracted 2--10~keV light curve. Two strong peaks are present in the plot, one
at $\sim$165~cycles~day$^{-1}$ ($525\pm1$~s) and the other at
$\sim$15~cycles~day$^{-1}$ ($5574\pm120$~s). The 2--10~keV light curve
folded at the shorter period is shown in Fig.~\ref{DD_Cir_folded}
where a coherent modulation is seen, which is nonetheless also consistent
with a constant flux. Unfortunately the target is too faint to construct meaningful 
light curves in different energy bands to investigate any variation in modulation depth 
with energy. 

The X-ray spectrum of this source is also too faint to obtain a meaningful model fit.
There is a pre-main sequence star in the \emph{RXTE} field of view, but its position
in the detector means it should contribute no more than 1/4 of the total count rate.

The possible modulation seen at 525~s means that DD Cir may be an IP, but a more 
sensitive detector is needed to prove the case either way.

\subsection{IGR J14536$-$5522}

J1453 was discovered as a hard X-ray source with \emph{INTEGRAL}
\citep{kuiper06}. Soon afterwards it was observed with the \emph{Swift} BAT
and found to be a relatively hard source with $kT\sim 30-40$~keV for a
single temperature Bremsstrahlung fit, exhibiting variability
\citep{mukai06}. Pointed \emph{Swift} XRT observations gave a flux of
$3.3 \times 10^{-11}$~ergs~cm$^{-2}$~s$^{-1}$ at 0.4--10~keV, and revealed a
complex spectrum requiring at least two components to fit the data
\citep{mukai06}. \citet{masetti06c} used the CTIO 1.5~m telescope to
get optical spectroscopy of J1453, on the basis of the Balmer and HeI
lines they classified it as a CV. Optical spectroscopy obtained from the 
SALT in 2006 exhibits a clear modulation at 3.1565(1)~hr, but SAAO 1.9~m
optical photometry does not \citep{potter10}. After these data were taken,
optical circular polarization from the source was discovered to be modulated at
the orbital period, which provided a strong indication that J1453 is a polar 
\citep{potter10}.

\subsubsection{Observations and results}

\emph{RXTE} data were collected over the course of just over one day (see
Table~\ref{observing_log}). The total good time on target
(31\,320~s) was split into 13 segments in PCU2. The raw target count
rate varied between 4.3 and 11.2 ct~s$^{-1}$ and the raw background
(generated from the calibration files) was between 2.7 and 4.1
ct~s$^{-1}$.

\begin{figure}
  \begin{center}
  \resizebox{\hsize}{!}{{\includegraphics{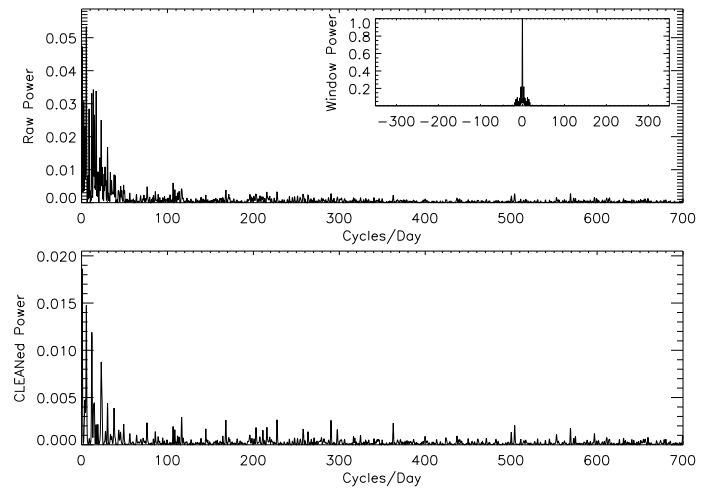}}}
  \caption{The {\sc clean}ed power spectrum of the 2--10~keV light curve of J1453. 
  The upper plot shows the raw power spectrum with the window function insert; the lower
  plot shows the deconvolved ({\sc clean}ed) power spectrum.}
  \label{J1453_cleaned}
  \end{center}
\end{figure}

\begin{figure}
  \begin{center}
  \resizebox{\hsize}{!}{{\includegraphics{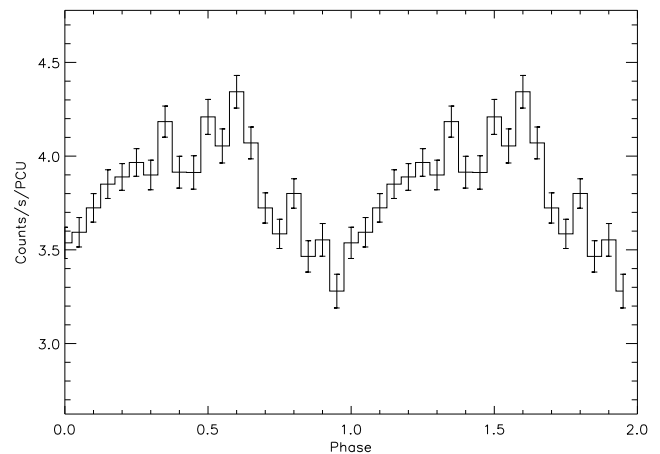}}}
  \caption{The 2--10~keV background subtracted light curve of J1453 folded at 3746~s, 
  with an arbitrary zero point. Two periods are shown for clarity.}
  \label{J1453_folded_at_3746}
  \end{center}
\end{figure}

When the background subtracted 2--10~keV light curve was analysed with 
the {\sc clean} algorithm, no evidence of any periodicity at the 3.1~hr 
(11\,363~s) spectroscopic period was seen. However, three periods are evident: 
$3746\pm110$~s, $7202\pm220$~s and $15\,594\pm1123$~s (see Fig.~\ref{J1453_cleaned}). 
These values are potentially a period and its first and third harmonics (within
uncertainties). Fig.~\ref{J1453_folded_at_3746} shows a plot of the data
folded at the 3746~s period. There is some 
indication of a trend in the modulation depth of the folded light curves
of each potential period to decrease with increasing energy (see
Table~\ref{J1453_modulation_depths}), but the trend is only marginally significant.

\begin{table*}
  \centering
  \caption{Modulation depths of the folded light curves of J1453, in different energy bands and at each of the potential periods.}
  \label{J1453_modulation_depths}
  \centering
  \begin{tabular}{l@{\hspace{1em}}r@{--}l l@{\hspace{2em}}cccc}
    \hline\hline
    \multicolumn{3}{c}{Energy band} & \multicolumn{4}{c}{Modulation depth}               & Fitted mean\\
    \multicolumn{3}{c}{(keV)}       & (\%)       & (\%)       & (\%)       & (\%)       & (ct~s$^{-1}$~PCU$^{-1}$) \\
    \multicolumn{3}{c}{}            & 3746       & 7202       & 15\,594    & 11\,363\\
    \hline
    &2&10                 & $8\pm1$    &   $10\pm1$ & $12\pm1$   & $4\pm1$   & 3.8\\
    &2&4                  & $11\pm1$   &   $11\pm1$ & $14\pm1$   & $7\pm1$    & 0.9\\
    &4&6                  & $6\pm1$    &   $11\pm1$ & $12\pm1$   & $4\pm1$    & 1.4\\
    &6&10                 & $8\pm1$    &   $9\pm1$  & $9\pm1$    & $3\pm1$    & 1.6\\
    &10&20            & $6\pm3$    &   $7\pm3$  & $7\pm1$    & $1\pm3$    & 0.7\\
    \hline\hline
  \end{tabular}
\end{table*}

\begin{figure}
  \begin{center} \resizebox{\hsize}{!}{\rotatebox{270}{\includegraphics{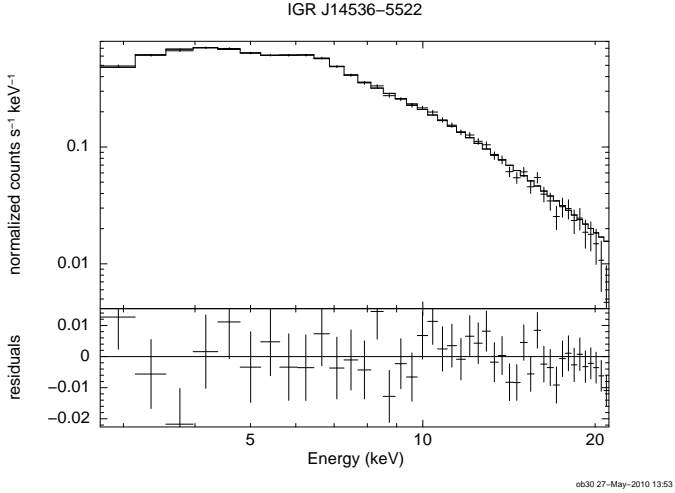}}}
  \caption{2.5--20~keV mean spectrum of J1453 fitted with a photoelectrically
  absorbed power law plus iron line profile.}
  \label{J1453_power_spectrum}
  \end{center}
\end{figure}

Fitting models to the J1453 X-ray spectrum required the column density be pegged to a
lower limit of the Galactic column density for the Bremsstrahlung
fit. The best fit was the power law fit, yielding a $\chi^2_{r}=0.9$
(see Fig.~\ref{J1453_power_spectrum} and also
Table~\ref{spectral_fits}).

\subsubsection{Discussion}

The peaks in the power spectrum of J1453 are not typical of spin periods
in IPs. The shortest period corresponding to the peaks seen is 3746$\pm$110~s, 
which would make it one of the longest spin periods of the IPs. This period is also not
seen in the optical data. As such the evidence for this as a spin
period is weak. The same is true for the 7202~s and 15\,594~s
peaks. The lack of any strong spin period candidates implies that
J1453 is probably not an IP.

In the absence of previous observations, these periods would be potential candidates 
for the orbital period of the system. In fact the reported orbital period 
(3.1565~hr $= 11360$~s) is close to what would be the second harmonic of 3746~s, 
with the two longer periods seen being further (sub-)harmonics. The uncertainties on 
these periods are relatively large however, and the correspondence may jut be a
coincidence. A longer base line in the X-ray data would constrain this
period further and conclusively show if this was real. 

The spectral profiles indicated a photoelectrically absorbed
power law plus iron line was the preferred fit, although the Bremsstrahlung fit 
was also acceptable, with a slightly higher $\chi^2_{r}$. Both fits are consistent with
a magnetic CV interpretation, with the clear detection of an iron emission line.

The flux seen here is significantly higher than previously measured --
4.1~$\times$10$^{-11}$~ergs~cm$^{-2}$~s$^{-1}$ in the 2--10~keV energy
band as opposed to  3.3~$\times$10$^{-11}$~ergs~cm$^{-2}$~s$^{-1}$ at
0.4--10~keV reported previously. There is one other X-ray source in the 
\emph{RXTE} field of view of J1453 with an estimated count rate of
0.96~ct~s$^{-1}$~PCU$^{-1}$. This other source has also 
been detected with \emph{INTEGRAL} and therefore is a hard X-ray source
too. This may affect the spectrum in a complex way that we cannot
characterise here.

\subsection{IGR J16167$-$4957}

IGR J16167$-$4957 (J1616) was discovered as a hard X-ray source in
the \emph{INTEGRAL}/IBIS survey \citep{barlow06}. Soon after this
\citet{tomsick06} constrained the position of J1616 with
\emph{Chandra} data and concluded it was not an HMXB. At the same
time \citet{masetti06} identified J1616 as a CV in their optical
spectroscopy campaign to identify \emph{INTEGRAL} objects.

\citet{pretorius09} reported a 5.004~hr (18\,014~s) orbital period seen in
optical spectroscopy, but no persistent coherent modulation in
high-speed photometry. A 585~s modulation was seen in some of
\citet{pretorius09} data, but not all, so is therefore likely a QPO.

\subsubsection{Observations and results}

J1616 was observed with \emph{RXTE} over two consecutive days (see
Table~\ref{observing_log}). The total good time was 34\,864~s in
PCU2. The 2--10~keV energy band raw count rate varied between 5.4--14.8
ct~s$^{-1}$, and the background rate was 2.8--4.1 ct~s$^{-1}$. The average
background subtracted count rate was 5.2 ct~s$^{-1}$.
The \emph{ROSAT} Bright Source Catalog has one other object in the \emph{RXTE} field of view at
approximately half a degree from J1616, which has twice the \emph{ROSAT} count
rate. This is likely to be a significant source of contamination in our 
observation. For this reason we also used archived \emph{XMM} data to better
constrain this source.

\begin{figure}
  \includegraphics[width=\hsize]{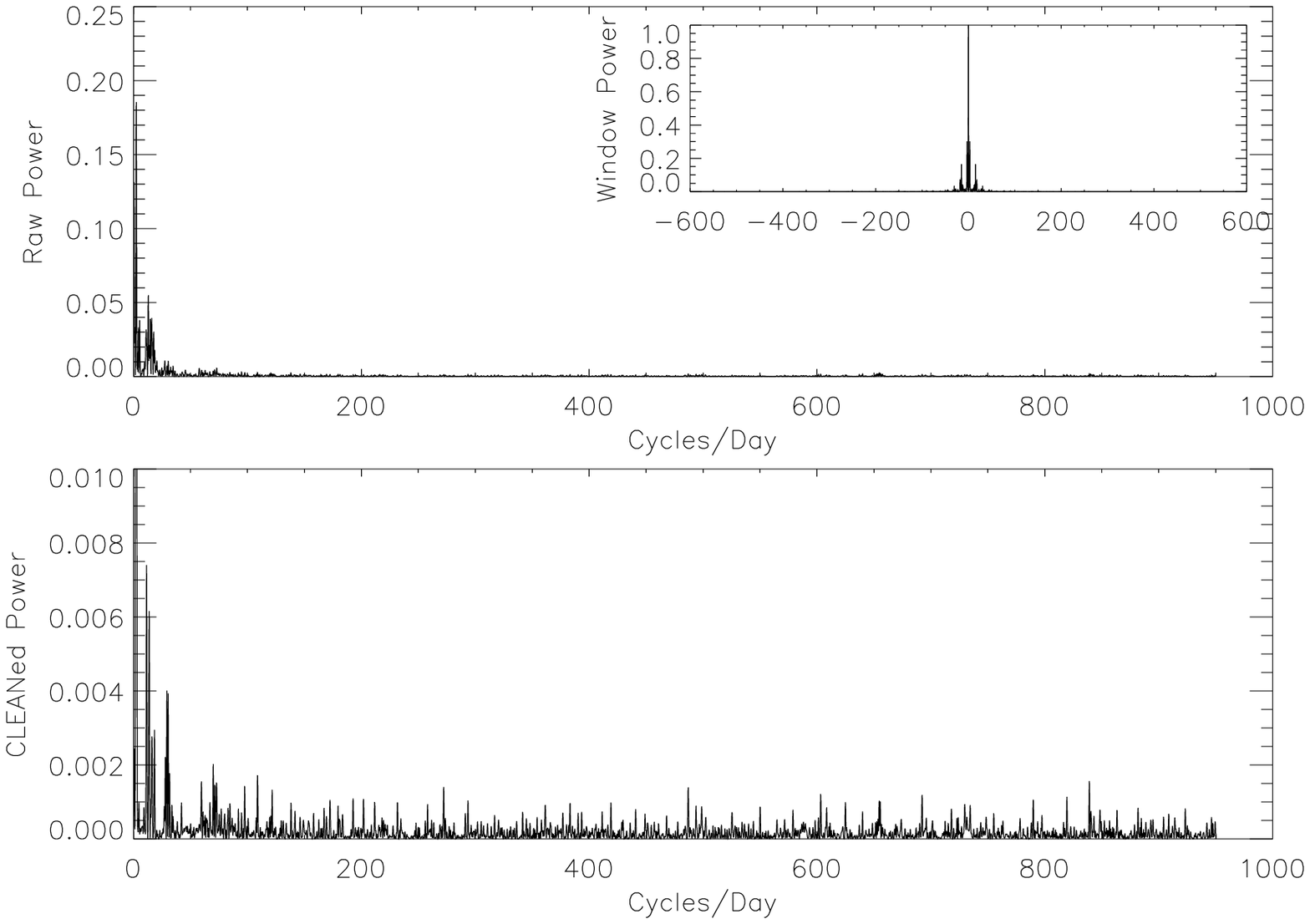}
  \caption{The {\sc clean}ed power spectrum of the 2--10~keV light curve of J1616. 
  The upper plot shows the raw power spectrum with the window function insert; the lower
  plot shows the deconvolved ({\sc clean}ed) power spectrum.}
  \label{J1616_cleaned}
\end{figure}

Fig.~\ref{J1616_cleaned} shows the {\sc clean}ed power spectrum of the 
2--10~keV \emph{RXTE} background subtracted light curve of J1616. There
are no significant peaks in the plot that are commensurate with a
potential spin period. This was also the case in the \emph{XMM}
data. The largest peak is at a period of 10.2$\pm$1.3~hr.

\begin{figure}
  \begin{center}
    \resizebox{\hsize}{!}{\rotatebox{-90}{\includegraphics{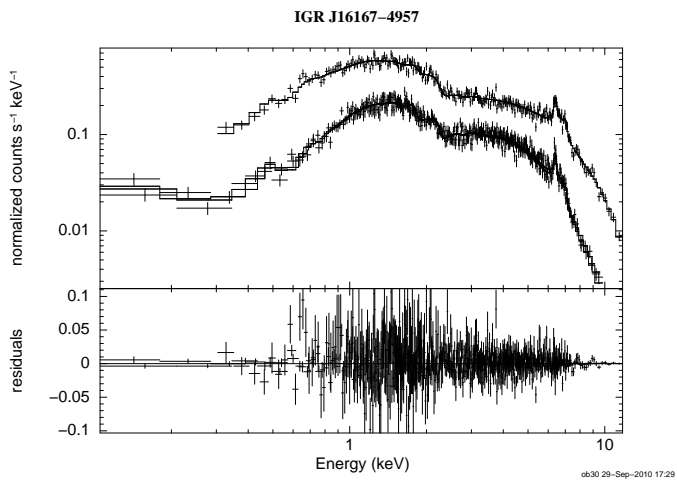}}}
    \caption{0.2--10 keV mean spectrum of J1616 fitted with two
      partial covering absorbers and a power law with iron line
      profiles. The upper data is the pn-data and the lower the two
      MOS data sets.}
    \label{J1616_xmm}
  \end{center}
\end{figure}

Fig.~\ref{J1616_xmm} shows the \emph{XMM}-EPIC spectrum of J1616 in the range
0.2--10~keV. A power law with two partial covering absorbers gave the
best fit to the data. Three gaussians were also included in the fit to
model the iron line features. Further to this an {\sc OVII} edge was added
at 0.7393~keV. The fit parameters are shown in
Table~\ref{J1616_spectral_fits_parameters}.

\renewcommand{\tabcolsep}{3pt}
\renewcommand{\arraystretch}{1.3}
\begin{table*}
  \centering
  \caption{Power law fit with two partial covering absorbers and three gaussians of J1616.}
  \label{J1616_spectral_fits_parameters}
  \centering
  \begin{tabular}{@{}llll@{}llllll@{}lll@{}llllll@{}}
    \hline\hline
    wabs $n_H$           & {\sc edge} $\tau$     & \multicolumn{5}{c}{Partial covering absorbers ($n_H$, Cv Frc)}                                & $\Gamma$             & \multicolumn{8}{c}{Iron lines (E, EW)}                                              & $\chi_r^2$ & \multicolumn{2}{c}{Flux}\\ \cline{3-7} \cline{9-16} \cline{18-19}
                         &                       & \multicolumn{2}{c}{1}                       & & \multicolumn{2}{c}{2}                         &                      & \multicolumn{2}{c}{1}      & & \multicolumn{2}{c}{2}      & & \multicolumn{2}{c}{3}     &            & pn & MOS\\ \cline{3-4} \cline{6-7} \cline{9-10} \cline{12-13} \cline{15-16}
    $10^{22}$ cm$^{-2}$   &                       & $10^{22}$ cm$^{-2}$  & --                    & & $10^{22}$ cm$^{-2}$    & --                    & --                   & keV                  & eV  & & keV                  & eV  & & keV                  & eV &            & \multicolumn{2}{c}{$10^{-11}$ ergs cm$^{-2}$ s$^{-1}$}\\
    \hline
    $0.27^{+0.02}_{-0.03}$ & $0.29^{+0.13}_{-0.13}$ & $2.55^{+1.31}_{-1.24}$ & $0.39^{+0.16}_{-0.11}$ & & $11.18^{+6.81}_{-3.57}$ & $0.43^{+0.09}_{-0.12}$ & $1.52^{+0.11}_{-0.11}$ & $6.40^{+0.02}_{-0.02}$ & 105 & & $6.65^{+0.10}_{-0.09}$ & 272 & & $7.00^{+0.09}_{-0.05}$ & 38 & 1.05       & 1.44 & 1.51\\
    \hline\hline
  \end{tabular}
\end{table*}

\subsubsection{Discussion}

There is no candidate period in the power spectrum that indicates a 
coherent modulation associated with a potential spin period and the 585~s
($\sim148$ cycles~day$^{-1}$) QPO seen by \citet{pretorius09} is not seen
here either. The 10.2~hr peak that we see is at roughly twice the orbital 
period reported by \citet{pretorius09}, however the uncertainty here is relatively 
large, so this may be coincidence.

The spectrum is fairly typical of IPs and suggests a power law with a
partial covering absorber with iron line features. Further to this,
the detection of the {\sc OVII} edge implies the existence of a warm
absorber - likely the pre-shock flow being photoionized. This has been
seen in two other IPs (V1223 Sgr \citep{mukai01} and 1RXS
J173021.5-055933 \citep{demartino08}) and further adds to the case for
IP classification. It should be noted
that two other models gave only a slightly worse fit - two partial
covering absorbers on an {\sc apec+apec} model and a power law distribution
of covering fractions ({\sc pwab}) on a {\sc mkcflow}. $\chi_r^2$= 1.07 and
1.06 respectively. 

The optical spectra taken by \citet{masetti06} and \citet{pretorius09}
along with the X-ray spectrum seen here point towards an IP
classification of J1616. However, the lack of any feasible spin period
candidate in either the optical or X-ray means that the classification cannot
be confirmed. Perhaps the geometry of the accretion
column is such that the magnetic and spin axis are aligned (or very
nearly aligned), so no spin modulation would be seen, as discussed
by \citet{ramsay08}. 

The IP classification of J1616 is thus unproven, though it remains as a 
candidate system.

\subsection{V2487 Oph}
V2487~Oph was discovered in the optical as a possible nova in 1998 at
magnitude 9.5 \citep{nakano98}.  \citet{lynch00} showed a plot of the
rapid decline in visual magnitude (courtesy of the AAVSO), indicating
V2487~Oph was a very fast nova. They also presented NIR spectra,
showing an overabundance in carbon soon after the
outburst. \citet{hachisu02} modelled the optical light curve and
concluded it has a WD mass of 1.35$\pm$0.01$M_{\odot}$, and the mass
transfer rate indicated a recurrence period of about 40
years. \citet{hernanz02} reported \emph{XMM} data of V2487~Oph taken
2.7 years after its discovery. They fit their spectrum with a
two-temperature plasma model ($T_{low}$ = 0.2 keV and $T_{high} \geq$
48 keV), which suggests a shocked gas (due an accretion flow) was
present. They also find an iron line at 6.4 keV. The error circles of
V2487~Oph and 1RXS J173200.0--191349 are coincident, and the \emph{XMM}
and \emph{ROSAT} fluxes are similar, indicating they may be the same
source. V2487~Oph was then found in the \emph{INTEGRAL/IBIS} survey
\citep{barlow06}. \citet{pagnotta09} searched archival data for
previous eruptions and found one occurred in 1900, confirming
V2487~Oph as a recurrent nova.

\subsubsection{Observations and results}
V2487 Oph was observed by \emph{RXTE} over two consecutive days (see
Table~\ref{observing_log}). The total good time was 42\,312~s in PCU2.
The 2--10~keV energy band raw count rate varied between
2.8--6.3~ct~s$^{-1}$, and the background count rate (generated from
the calibration files) 0--2.8~ct~s$^{-1}$. The average background
subtracted count rate was 1.1~ct~s$^{-1}$. There is another X-ray source in
the field of view that we would expect to contribute a comparable count rate
to V2487~Oph.

\begin{figure}
  \includegraphics[width=\hsize]{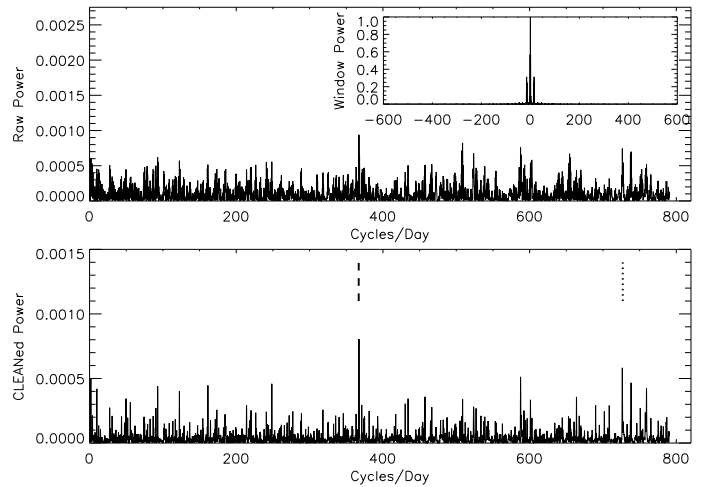}
  \caption{The {\sc clean}ed power spectrum of the 2--10~keV light curve of V2487~Oph. 
  The upper plot shows the raw power spectrum with the window function insert; the lower
  plot shows the deconvolved ({\sc clean}ed) power spectrum. The
  dashed line is at 367 cycles~day$^{-1}$, the dotted line at 727
  cycles~day$^{-1}$.}
  \label{V2487_cleaned}
\end{figure}

Fig.~\ref{V2487_cleaned} shows the {\sc clean}ed power spectrum of the 
background subtracted 2--10~keV light curve. The largest peak is at
approximately 367~cycles~day$^{-1}$ (235.2$\pm$0.1~s), the second largest peak
is at 118.9$\pm$0.1~s - very close (but not within errors) of being
a potential first harmonic of the 235.2 peak. The folded data (in all
bands) shows only marginally coherent modulation, and as such we do
not consider the 235.2~s peak to be a secure period, merely a
candidate.

\begin{figure}
  \begin{center}
    \resizebox{\hsize}{!}{\rotatebox{-90}{\includegraphics{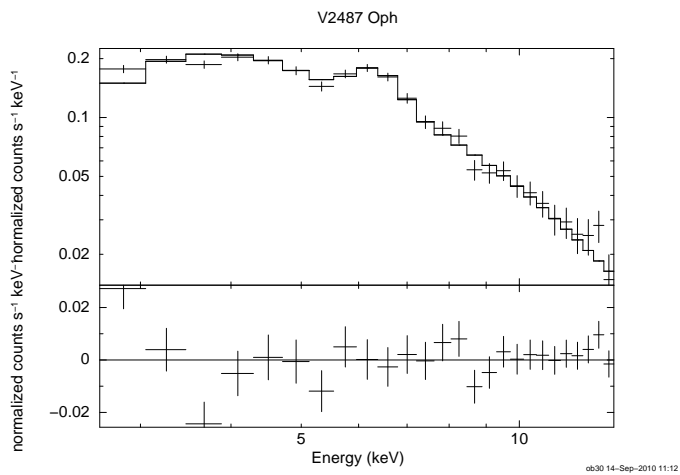}}}
    \caption{2.5--12 keV mean spectrum of V2487 Oph fitted with a photoelectrically
      absorbed power law plus iron line profile.}
    \label{V2487_power}
  \end{center}
\end{figure}

The mean X-ray spectrum was fitted with both a photoelectrically absorbed
bremsstrahlung and an absorbed power law, with an iron emission line in each case, 
over the range 2.5--12~keV. The results are summarised in
Table~\ref{spectral_fits}. The column density was pegged to a lower limit of $2\times 10^{21}$~cm$^{-2}$
(D\&L). Fig.~\ref{V2487_power} shows the power law fit. The
2--10~keV flux of the fit was 1.2$\times 10^{-11}$~ergs cm$^{-2}$ s$^{-1}$.

\subsubsection{Discussion}
The lack of any significant coherent modulation is unsurprising given
the recent nova eruption. It may be some time before the system
settles down to a state where modulation can be seen from the
accretion column (if this ever does happen). The spectrum is indicative of an IP, with a strong
iron line. There is however an excess at lower energies which skews
the model fits, this may be from residual nuclear burning after the
nova, or from the more traditional soft X-ray source in IPs. The flux
is approximately three times higher
than that in \citet{hernanz02}, but this is in a sightly different
energy band to theirs (0.3--8~keV), and as noted earlier there is
another source in the field of view. 

V2487~Oph therefore remains as a candidate IP until it reaches a more
stable state at which point the presence of periodicities can be
probed further.

\subsection{V2069 Cyg}

V2069~Cyg (RX~J2123.7+4217) was identified as a CV by \citet{motch96} in the
\emph{ROSAT} all-sky survey and \citet{thorstensen01} measured an orbital period
of 0.311683(2)~days (7.480~hr) from optical
spectroscopy. \citet{barlow06} subsequently found V2069~Cyg in their 
\emph{INTEGRAL}/IBIS survey. \citet{demartino09} reported preliminary analysis 
of \emph{XMM} data of V2069~Cyg in an ATel, where they find a fundamental frequency 
of 116.3 cycles~day$^{-1}$ and its harmonics, giving a period of 743.2$\pm$0.4~s. They
also give a fit to the spectrum of a 16~keV thermal plasma with a 56~eV blackbody 
component and a Gaussian at 6.4~keV, being absorbed by a partially covering model.

\subsubsection{Observations and results}

V2069~Cyg was observed with \emph{RXTE} over two consecutive days (see
Table~\ref{observing_log}). The total good time was 37\,472~s in
PCU2. The 2--10~keV energy band raw count rate varied between
2.6--6.2~ct~s$^{-1}$, and the background count rate (generated from
the calibration files) was 0--4.2~ct~s$^{-1}$. The average background
subtracted count rate was 1.1~ct~s$^{-1}$. 
There is one other source in the {\em RXTE} field of view of V2069 Cyg, but it is at
the very edge of the detector, and is also very faint, and so is
ignored here.

\begin{figure}
  \includegraphics[width=\hsize]{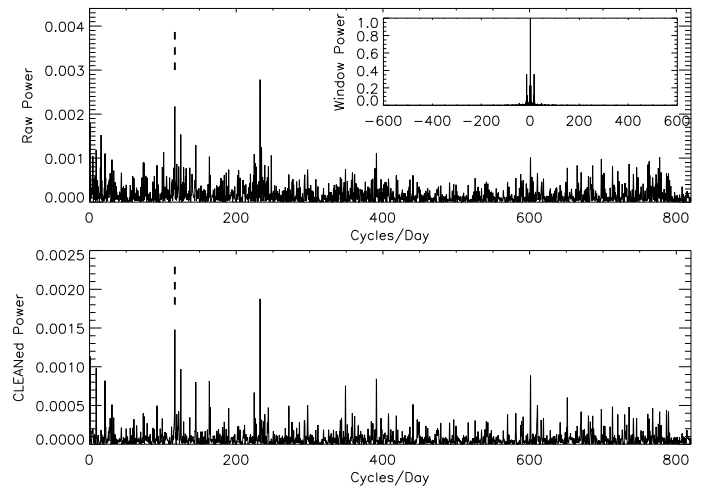}
  \caption{The {\sc clean}ed power spectrum of the 2--10~keV light curve of V2069~Cyg. 
  The upper plot shows the raw power spectrum with the window function insert; the lower
  plot shows the deconvolved ({\sc clean}ed) power spectrum. The
  dashed line is at 116 cycles~day\textsuperscript{-1} (743~s).}
  \label{V2069_cleaned}
\end{figure}

Fig.~\ref{V2069_cleaned} shows the {\sc clean}ed power spectrum of the 
background subtracted 2--10~keV light curve. Peaks are
seen at approximately 115~cycles~day$^{-1}$, 230~cycles~day$^{-1}$ and
350~cycles~day$^{-1}$. If these correspond to the fundamental and first and
second harmonics, then this implies a period of 743.2$\pm$0.9~s.

\begin{figure}
  \includegraphics[width=\hsize]{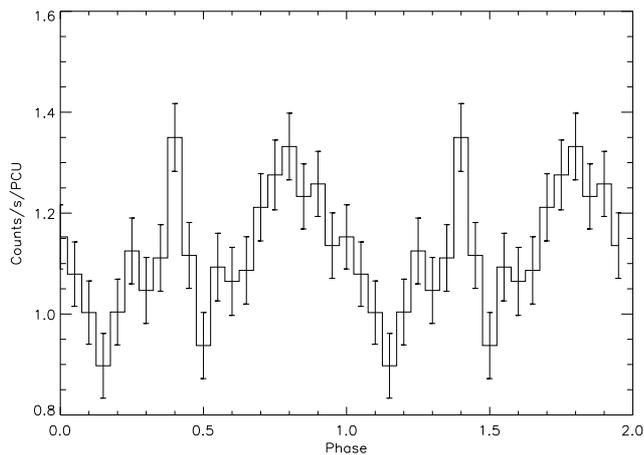}
  \caption{The 2--10~keV light curve of V2069 Cyg folded at 743.2~s
    with an arbitrary zero point. Two periods are shown.}
  \label{V2069_2_10_folded_743p2}
\end{figure}

The 2--10~keV light curve folded at 743.2~s is shown in
Fig.~\ref{V2069_2_10_folded_743p2}. A clear double peak modulation
is present -- as expected from the harmonic structure seen in the power spectrum. 
Folding the light curve in each energy band at this period and analysing the modulation 
depth shows that there is a trend of an increasing modulation depth with
decreasing energy (see Table~\ref{V2069_modulation_depth}).
There is no coherent modulation in the light curves when folded at the
7.48~hr spectroscopic orbital period of \citet{thorstensen01}. 

\begin{table}
\centering
\caption{Modulation depths of the V2069 Cyg X-ray light curve folded at 743.2~s.}
\label{V2069_modulation_depth}
\begin{tabular}{lll}
\hline
\hline
Energy band & Mean        & Modulation depth\\
(keV)       & ct s$^{-1}$ & (\%)\\
\hline
2--10       & 1.13$\pm$0.02 &  7.9$\pm$2.5\\
2--4        & 0.26$\pm$0.01 & 12.7$\pm$3.9\\
4--6        & 0.39$\pm$0.01 &  8.0$\pm$3.0\\
6--10       & 0.48$\pm$0.01 &  5.8$\pm$2.7\\
10--20      & 0.24$\pm$0.01 &  1.4$\pm$6.2\\
\hline
\hline
\end{tabular}
\end{table}

\begin{figure}
  \begin{center}
    \resizebox{\hsize}{!}{\rotatebox{-90}{\includegraphics{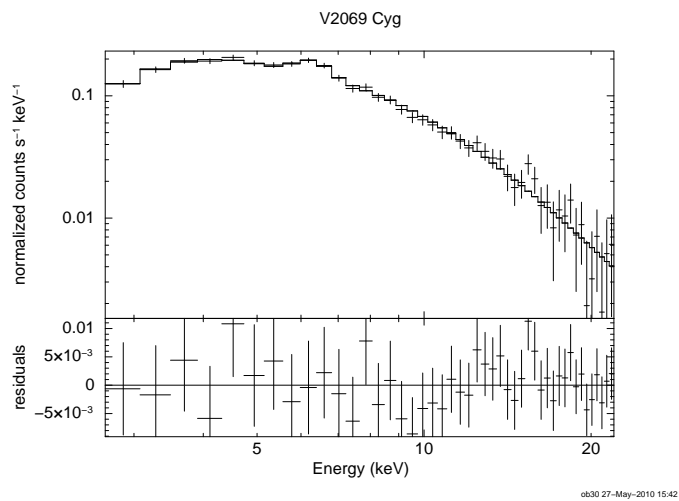}}}
    \caption{2.5--20 keV mean spectrum of V2069 Cyg fitted with a photoelectrically
      absorbed Bremsstrahlung plus iron line profile.}
    \label{V2069_brems}
  \end{center}
\end{figure}

The mean X-ray spectrum was fitted with both a photoelectrically absorbed
bremsstrahlung and an absorbed power law, with an iron emission line in each case, 
over the range 2.5--20~keV. The results are summarised in
Table~\ref{spectral_fits}. The column density was pegged to a lower limit of $3.3\times 10^{21}$~cm$^{-2}$
(D\&L). Fig.~\ref{V2069_brems} shows the bremsstrahlung fit. The
2--10~keV flux of the fit was 1.2$\times 10^{-11}$~ergs cm$^{-2}$ s$^{-1}$.

\subsubsection{Discussion}

We interpret the period seen here (743.2$\pm$0.9~s) as the spin period
of the white dwarf in V2069~Cyg. This is in very good agreement with the
preliminary analysis of \citet{demartino09} who find a 743.2$\pm$0.4~s
period. Furthermore, the trend of decreasing modulation depth with increasing X-ray 
energy is similar to that commonly seen in other IPs, and indicates that photoelectric
absorption is a major contributor to the modulation. The hard X-ray
spectrum seen here, with an iron line at 6.4~keV is typical of IPs.

Given the spin period modulation, the trend in the modulation depth
and the spectral fit, V2069~Cyg is therefore confirmed as an
intermediate polar.

\section{Conclusion}

V2069 Cyg is conclusively proven here to be an intermediate polar with a white 
dwarf spin period of 743.2~s and a characteristic hard X-ray spectrum with an iron line. 
Our data are consistent with the previous claims for J0023 to be an intermediate polar, 
but do not add further to the information about the object. Likewise, our data are 
consistent with the polar interpretation of J1453, revealing hard X-ray signals that 
are likely to be harmonically related to the previously detected orbital period. 
The \emph{RXTE} observations of J0153, J0162, V436 Car and DD Cir are each essentially too 
faint to allow further conclusions to be drawn. Although we see tentative evidence for 
a 292~s pulsation in J0162 and a 525~s pulsation in DD Cir, no coherent X-ray pulsations 
are conclusively detected in any of these four objects.
Finally, our observations of J1616 and V2487~Oph reveal X-ray spectra that are 
characteristic of an IP, but no coherent X-ray pulsations are seen, so these objects 
remain candidate IPs. We speculate that perhaps these systems may have
their magnetic axis aligned with their spin axis. This could account
for the lack of significant coherent spin modulation while still allowing
the spectral properties seen here.

\emph{INTEGRAL} found several IPs in its initial survey and this was largely a surprise
to the CV community. Since then it has found many more as well as finding many 
candidate IPs. Follow up of these candidates with ground based optical telescopes 
has revealed periods in some cases, and spectra in most cases that are commensurate 
with an IP classification. X-ray classification is an integral part of this
process, and the work done in this campaign with \emph{RXTE}, and that done
by others with \emph{XMM} has been invaluable. With the follow-up of all 
IP candidates discovered by \emph{INTEGRAL} now nearing completion, it may be that we
have now found virtually all the hard X-ray selected IPs that are bright enough to observe 
with current instrumentation.

\bibliographystyle{aa}
\bibliography{ref}

\end{document}